\documentclass[aps]{revtex4}
\usepackage{bm}
\usepackage{epsfig,amsmath,graphicx,amssymb}
\def\be{\begin{equation}}
\def\ee{\end{equation}}
\def\bea{\begin{eqnarray}}
\def\eea{\end{eqnarray}}
\begin{document}
\title{THE FIRST AND SECOND VARIATION OF THE TOTAL ENERGY OF CLOSED DUPLEX DNA IN PLANAR CASE
 }
\author{Xiao-hua Zhou}
\email{xhzhou08@gmail.com;zhouxh1980@gmail.com}
\affiliation{Department of Mathematics and Physics, The Fourth
Military Medical University, Xi'an 710032, People's Republic of
China }

\date{\today}

\begin{abstract}

DNA's shape mostly lies on its total energy $F$. Its corresponding
equilibrium shape equations can be got by classical variation
method: letting the first energy variation $\delta^{(1)}F=0$. Here
we not only provide the first variation $\delta^{(1)}F$ but also
give the second variation $\delta^{(2)}F$ in planar case. Moreover,
the general shape equations of DNA are obtained and a mistake in
[Zhang, \emph{et al}. {\it Phys. Rev. E} {\bf 70} 051902 (2004)] is
pointed out.

\end{abstract}

\maketitle


\section*{1. Introduction}

Closed duplex DNA molecules often have complex configurations, such
as supercoil, and those structures play an important role in gene
regulation.$^{1}$ Theoretical analysis about those configurations
are based on the elastic rod theory. In 1859, Kirchhoff given the
equilibrium shape equations of thin elastic rod and his methods has
been generally used to describe the holistic conformations of duplex
DNA.$^{2-5}$ Benham$^{2}$ treated a homogeneous isotropic rod of the
DNA by Kirchhoff analogy and gave some possible shapes. As for the
closed duplex DNA molecules, its equilibrium equations are obtained
$^{6}$, and considerable results are attained.$^{7-11}$

Following the development of modern experimental techniques, such as
optical tweezers,$^{12}$ micromanipulation,$^{13-15}$ and other
techniques,$^{16-18}$ researchers are presented with more
opportunities to probe into the microstructure of individual great
molecule than ever before. The ever growing volume of experimental
data provides us with the probabilities to improve the existing
theoretical models.

Recently, considering the microcosmic configurations of DNA, more
practical models, such as the anisotropic model$^{19}$ are put
forward, and new energy terms are introduced,$^{20-24}$ and new
methods, such as Monte Carlo simulation$^{20,25}$ and path
integral,$^{21}$ are adopted. Due to those improved theories, some
theoretical results are highly consistent with the experimental
values.$^{21,24}$

 The geometrical configuration of duplex DNA is
shown in Fig.1. The axis of the molecule can be written as
$\textbf{R}(s)$ parameterized by the arclength $s$. The total energy
of a simple elastic model for the closed duplex DNA can be written
as $^{9}$
\be
F=\oint\left(\frac{A}{2}K^{2}+B\tau\right)ds,
\ee
where $K=K(s)$ and $\tau=\tau(s)$ are the curvature and torsion of
the axis $\textbf{R}(s)$. $A$ and $B$ are the bending rigidity and
torsional rigidity, respectively. Some researchers also choose a
simpler model without the torsional term.$^{26}$

\begin{figure}\centering
\includegraphics[scale =0.3]{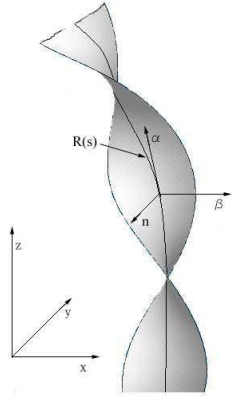}
\caption{The geometrical configuration of duplex DNA molecule. The
central line of the ribbon follows the molecular axis with
coordinates $\textbf{R}(s)$ parameterized by the arclength $s$. A
nature coordinates is defined: the tangent vector
$\boldsymbol{\alpha}$, the main normal vector $\textbf{n}$ and the
vice-normal vector $\boldsymbol{\beta}$}
\end{figure}

This letter is composed as follows. In Sec.~\textbf{2}, we will get
the equilibrium shape equations of closed duplex DNA by variation
theory in its nature coordinates system. In Sec.~\textbf{3}, a
general model is considered and its equilibrium shape equations are
attained. In Sec.~\textbf{4}, we give the second variation
$\delta^{(2)}F$ in planar case. Finally, a short discussion is given
in Sec.~\textbf{5}.

\section*{2. Classical Equilibrium Shape Equations of Closed Duplex DNA}

For a more general case, we choose the total energy of DNA
\be
F=\oint\left[\frac{A}{2}(K-C_0)^{2}+B(\tau-T_0)\right]ds+\lambda\oint
ds,
\ee
where $C_0$ and $T_0$ are two constants, and we call them
spontaneous curvature and spontaneous tension respectively, and
$\lambda$ is a Lagrangian multiplier which can be taken as the
tensile rigidity. Fig. 1 shows a fragment of $\textbf{R}(s)$ in the
orthogonal coordinates system ($x,y,z$). Meantime, a nature (local)
coordinates can be described as
\be
(\boldsymbol{\alpha}, \textbf{n}, \boldsymbol{\beta}).
\ee
Where, $\boldsymbol{\alpha}=\dot{\textbf{R}}$ is the tangent vector
(an overdot denotes differential with respect to $s$),
$\textbf{n}=\ddot{\textbf{R}}/K$ is the main normal vector and $
\boldsymbol{\beta}=\boldsymbol{\alpha}\times\textbf{n}$ is the
vice-normal vector. Between those unit vectors and $K$ and $\tau$,
there are the Frenet formulaes
\bea
 & & \dot{\boldsymbol{\alpha}}=K(s)\textbf{n},\\
 & & \dot{\textbf{n}}=-K(s)\boldsymbol{\alpha}+\tau(s) \boldsymbol{\beta},\\
 & & \dot{ \boldsymbol{\beta}}=-\tau(s)\textbf{n}.
\eea
The equilibrium shape equations will be extracted by analyzing the
first variation
\be
\delta^{(1)}{F}=0.
\ee
Letting $g=\textbf{R}_{,x}\cdot\textbf{R}_{,x}$ ($
\textbf{R}_{,x}=\frac{d\textbf{R}}{dx}$), we have $ds=\sqrt{g}dx$.
So, Eq. (7) induced
\bea
\nonumber        \oint\bigg\{\left[A(K-C_0)\delta^{(1)}K+B\delta^{(1)}\tau\right]\sqrt{g} & & \\
  +\left[\frac{A}{2}(K-C_0)^{2}+B(\tau-T_0)+\lambda\right]\delta^{(1)}(\sqrt{g})\bigg\}dx &=& 0.
\eea
Before we get the equilibrium shape equations, some useful identical
equations should be calculated (see appendix A).

The $\textbf{R}(s)$ under small perturbations can be written as
\be
\textbf{R}^{'}(s)=\textbf{R}(s)+\psi(s)\textbf{n}+\varphi(s)\boldsymbol{\beta},
\ee
where $\psi(s)$ and $\varphi(s)$ are two small smooth functions. We
get
\bea
 & & \delta\textbf{R}=\textbf{R}^{'}-\textbf{R}=\psi\textbf{n}+\varphi\boldsymbol{\beta}\\
 & & \delta\textbf{R}_{,x}=(\delta\textbf{R})_{,x}=\sqrt{g}(\delta\textbf{R})_{,s}=\sqrt{g}(
     \dot{\psi}\textbf{n}+\psi\dot{\textbf{n}}+\dot{\varphi}\boldsymbol{\beta}+\varphi\dot{\boldsymbol{\beta}})
\eea
and
\bea
\nonumber
  \delta{g} &=& (\textbf{R}_{,x}+\delta{\textbf{R}_{,x}})\cdot(\textbf{R}_{,x}+\delta{\textbf{R}_{,x}})-\delta{\textbf{R}_{,x}}\cdot\delta{\textbf{R}_{,x}}\\\nonumber
            &=& 2\textbf{R}_{,x}\cdot\delta\textbf{R}_{,x}+\delta\textbf{R}_{,x}\cdot\delta\textbf{R}_{,x}=g[2\dot{\textbf{R}}\cdot(\delta\textbf{R})_{,s}
                +(\delta\textbf{R})_{,s}\cdot(\delta\textbf{R})_{,s}]\\
            &=& -2gK\psi+g\left[\dot{\psi}^2+\psi^2(K^2+\tau^2)+\dot{\varphi}^2+\varphi^2\tau^2+2(\psi\dot{\varphi}-\dot{\psi}\varphi)\tau\right].
\eea
Consequently, we have
\bea
\nonumber
 \delta{g^{\frac{1}{2}}} &=& (g+\delta g)^{\frac{1}{2}}-g^{\frac{1}{2}}=g^{\frac{1}{2}}\left[\frac{1}{2}\left(\frac{\delta g}{g}\right)-\frac{1}{8}\left(\frac{\delta g}{g}\right)^2+\cdots\right]\\
                         &=& -g^{\frac{1}{2}}K\psi+\frac{1}{2}g^{\frac{1}{2}}\left[\dot{\psi}^2+(\psi^2+\varphi^2)\tau^2+\dot{\varphi}^2+2(\psi\dot{\varphi}-\dot{\psi}\varphi)\tau\right]+\emph{O(3)},
\eea
\bea
\nonumber
 \delta{g^{\frac{-1}{2}}} &=& (g+\delta g)^{\frac{-1}{2}}-g^{\frac{-1}{2}}=g^{\frac{-1}{2}}\left[-\frac{1}{2}\left(\frac{\delta g}{g}\right)+\frac{3}{8}\left(\frac{\delta g}{g}\right)^2+\cdots\right]\\\nonumber
                          &=& g^{\frac{-1}{2}}K\psi-\frac{1}{2}g^{\frac{-1}{2}}\left[\dot{\psi}^2+(\psi^2+\varphi^2)\tau^2+\dot{\varphi}^2-2\psi^2K^2+2(\psi\dot{\varphi}-\dot{\psi}\varphi)\tau\right]\\
                          & & +\emph{O(3)}.
\eea
Where $\emph{O(3)}$ means the third and higher orders of $\psi$ and
$\varphi$. We can also attain
\bea
\nonumber
 \delta{\dot{\textbf{R}}} &=& \delta(g^{\frac{-1}{2}}\textbf{R}_{,x})=(g^{\frac{-1}{2}}+\delta
                               g^{\frac{-1}{2}})(\textbf{R}_{,x}+\delta\textbf{R}_{,x})-\dot{\textbf{R}}\\
                          &=& (1+g^{\frac{1}{2}}\delta g^{\frac{-1}{2}})(\delta\textbf{R})_{,s}+g^{\frac{1}{2}}\dot{\textbf{R}}\delta g^{\frac{-1}{2}}
\eea
This conclusion can be generalized to an arbitrate function
$\textbf{V}=\textbf{V}(s)$, which satisfies
\be
 \delta{\dot{\textbf{V}}}= (1+g^{\frac{1}{2}}\delta g^{\frac{-1}{2}})(\delta\textbf{V})_{,s}+g^{\frac{1}{2}}\dot{\textbf{V}}\delta g^{\frac{-1}{2}}
\ee
Submitting (11) and (14) into (15), we get
\bea
\nonumber
 \delta{\dot{\textbf{R}}} &=& K\psi\dot{\textbf{R}}+
                              \dot{\psi}\textbf{n}+\psi\dot{\textbf{n}}+\dot{\varphi}\boldsymbol{\beta}+\varphi\dot{\boldsymbol{\beta}}+K\psi(
                              \dot{\psi}\textbf{n}+\psi\dot{\textbf{n}}+\dot{\varphi}\boldsymbol{\beta}+\varphi\dot{\boldsymbol{\beta}})\\
                          & & -\frac{1}{2}\dot{\textbf{R}}\left[\dot{\psi}^2+\dot{\varphi}^2+(\psi^2+\varphi^2)\tau^2-2\psi^2K^2+2(\psi\dot{\varphi}-\dot{\psi}\varphi)\tau\right]+\emph{O(3)}.
\eea
Using (16), we have
\be
 \delta{\ddot{\textbf{R}}}= (1+g^{\frac{1}{2}}\delta g^{\frac{-1}{2}})(\delta\dot{\textbf{R}})_{,s}+g^{\frac{1}{2}} \ddot{\textbf{R}}\delta
 g^{\frac{-1}{2}}.
\ee
Insetting (14) and (17) into (18) we get
\bea
\nonumber
 \delta{\ddot{\textbf{R}}} &=& 2K\psi\ddot{\textbf{R}}+\dot{\textbf{R}}(\dot{K}\psi+K\dot{\psi})+\ddot{\psi}\textbf{n}+\psi\ddot{\textbf{n}}+2\dot{\psi}\dot{\textbf{n}}
                               +\ddot{\varphi} \boldsymbol{\beta}+\varphi\ddot{ \boldsymbol{\beta}}+2\dot{\varphi}\dot{
                               \boldsymbol{\beta}}\\\nonumber
                           & & +K\psi\left[2\ddot{\psi}\textbf{n}+2\psi\ddot{\textbf{n}}+2\varphi\ddot{\boldsymbol{\beta}}+2\ddot{\varphi}\boldsymbol{\beta}+4\dot{\psi}\dot{\textbf{n}}
                               +4\dot{\varphi}\dot{\boldsymbol{\beta}}+K\psi\ddot{\textbf{R}}+(\dot{K}\psi+K\dot{\psi})\dot{\textbf{R}}\right]\\\nonumber
                           & & -\ddot{\textbf{R}}\left[\dot{\psi}^2+\dot{\varphi}^2+(\psi^2+\varphi^2)\tau^2-2K^2\psi^2+2(\psi\dot{\varphi}-\dot{\psi}\varphi)\tau\right]\\\nonumber
                           & & +(\dot{K}\psi+K\dot{\psi})(\dot{\psi}\textbf{n}+\psi\dot{\textbf{n}}+\dot{\varphi}\boldsymbol{\beta}+\varphi\dot{\boldsymbol{\beta}})\\\nonumber
                           & & -\dot{\textbf{R}}\left[\tau\dot{\tau}(\psi^2+\varphi^2)-2K\dot{K}\psi^2-2K^2\psi\dot{\psi}+\dot{\tau}(\psi\dot{\varphi}-\dot{\psi}\varphi)+\tau^2(\psi\dot{\psi}+\varphi\dot{\varphi})\right.\\
                           & & \left.+\dot{\psi}\ddot{\psi}+\dot{\varphi}\ddot{\varphi}+\tau(\psi\ddot{\varphi}-\ddot{\psi}\varphi)\right]+\emph{O(3)}.
\eea
 Noting $K^2=\ddot{\textbf{R}}\cdot\ddot{\textbf{R}}$, we have
\be
\delta^{(1)}{K}=(\ddot{\textbf{R}}\cdot\delta^{(1)}\ddot{\textbf{R}})/K.
\ee
Submitting (19) into (20) and using the identical equations in
appendix A, we attain
\be
\delta^{(1)}{K}=\left(K^2-\tau^2\right)\psi+\ddot{\psi}-\dot{\tau}\varphi-2\tau\dot{\varphi}.
\ee
Now, we calculate
\bea
\nonumber \delta^{(1)}{\dot{\textbf{n}}} &=&
                                             \dot{\textbf{n}}\sqrt{g}\delta^{(1)}{(g^{-1/2})}+g^{-1/2}\left[\delta^{(1)}\left(\frac{\ddot{\textbf{R}}}{K}\right)\right]_{,x}\\
                                         &=& \left[\frac{K\delta^{(1)}{\ddot{\textbf{R}}}
                                             -\ddot{\textbf{R}}\delta^{(1)}{K}}{K^2}\right]_{,s}+\dot{\textbf{n}}K\psi.
\eea
Submitting (19) and (21) into (22), we get
$\delta^{(1)}{\dot{\textbf{n}}}$. Then, using the identic equations
in appendix A, the following useful term can be attained
\bea
\nonumber\dot{\textbf{n}}\cdot\delta^{(1)}{\dot{\textbf{n}}}  &=&
                                                            K^3\psi+K\left(\tau^2\psi-\dot{\tau}\varphi-\tau\dot{\varphi}+\ddot{\psi}\right)\\\nonumber                                            & & +K^{-1}\tau\left[\ddot{\tau}\psi-\tau^2\dot{\varphi}
                                                            +3\dot{\tau}\dot{\psi}+2\tau(\ddot{\psi}-\dot{\tau}\varphi)+\varphi^{(3)}\right]\\
                                                        & & +K^{-2}\dot{K}\tau\left(\tau^2\varphi-\dot{\tau}\psi-2\tau\dot{\psi}-\ddot{\varphi}\right).
\eea
Considering $\dot{\textbf{n}}\cdot\dot{\textbf{n}}=K^2+\tau^2$, we
have
\be
\delta^{(1)}{\tau}=\frac{\dot{\textbf{n}}\cdot\delta^{(1)}\dot{\textbf{n}}-K\delta^{(1)}{K}}{\tau}.
\ee
Taking (21) and (23) into (24), we get
\bea
\nonumber\delta^{(1)}{\tau} &=& K(2\tau\psi+\dot{\varphi})
                                +K^{-2}\dot{K}\left(\tau^2\varphi-\dot{\tau}\psi-2\tau\dot{\psi}-\ddot{\varphi}\right)\\
                            & & +K^{-1}\left[\ddot{\tau}\psi+3\dot{\tau}\dot{\psi}-\tau^2\dot{\varphi}+
                                 \varphi^{(3)}+2\tau\left(\ddot{\psi}-\varphi\dot{\tau}\right)\right].
\eea
Submitting (13), (21) and (25) into (8), we attain
\bea
\nonumber\delta^{(1)}{F} &=&
                             \oint\bigg\{-K\psi\left[\frac{A}{2}(K-C_0)^{2}+B(\tau-T_0)+\lambda\right]\\\nonumber
                         & & +A(K-C_0)\left[\left(K^2-\tau^2\right)\psi
                             -\dot{\tau}\varphi-2\tau\dot{\varphi}+\ddot{\psi}\right]\\\nonumber
                         & & +BK(2\tau\psi+\dot{\varphi})+BK^{-2}\dot{K}\left(\tau^2\varphi-\dot{\tau}\psi-2\tau\dot{\psi}-\ddot{\varphi}\right)\\
                         & & +BK^{-1}\left[\ddot{\tau}\psi+3\dot{\tau}\dot{\psi}-\tau^2\dot{\varphi}+\varphi^{(3)}+2\tau\left(\ddot{\psi}-\dot{\tau}\varphi\right)\right]\bigg\}
ds.
\eea
If perturbation is only on the main normal direction ($\varphi=0$),
we have
\bea
\nonumber F_0 &=& \frac{1}{2}A\left(K^3+2C_0\tau^2-K
                  C_0^2-2K\tau^2\right)-K\left(\lambda-B\tau-BT_0\right)\\\nonumber
              & & +BK^{-2}(K\ddot{\tau}-\dot{K}\dot{\tau}),\\\nonumber
          F_1 &=& BK^{-2}(3K\dot{\tau}-2\dot{K}\tau),\\\nonumber
          F_2 &=& A(K-C_0)+2BK^{-1}\tau,\\
                   \delta^{(1)}{F} &=& \oint\left(F_0\psi+F_1\dot{\psi}+F_2\ddot{\psi}\right)ds.
\eea
For two smooth functions $Q=Q(s)$ and $P=P(s)$, there is
\be
\oint Q^{(n)}P ds=(-1)^n \oint P^{(n)}Q ds,
\ee
where $n=1,2,3\cdots$. Using (28), (27) can be changed into
\be
\delta^{(1)} F=\oint
\left[\frac{1}{2}AK^3-\frac{1}{2}K\left(AC_0^2+2\lambda-2BT_0-2B\tau+2A\tau^2\right)+A\left(C_0\tau^2+\ddot{K}\right)\right]\psi
ds.
\ee
Because $\psi$ is an arbitrary function, so the equilibrium shape
equation under main normal perturbation is
\be
AK^3-K(AC_0^2+2\lambda-2BT_0-2B\tau+2A\tau^2)+2A(C_0\tau^2+\ddot{K})=0.
\ee
Another equilibrium shape equation under vice-normal perturbation
can be obtained through the former way. Choosing $\psi=0$, (26) can
be simplified as
\bea
\nonumber
     F_0 &=& BK^{-2}\dot{K}\tau^2+2BK^{-1}\dot{\tau}\tau-A(K-C_0)\dot{\tau},\\\nonumber
     F_1 &=& K(B-2A\tau)+2AC_0\tau-BK^{-1}\tau^2,\\\nonumber
     F_2 &=& -BK^{-2}\dot{K},\\\nonumber
     F_3 &=& BK^{-1},\\
\delta^{(1)} F &=&
\oint\left(F_0\varphi+F_1\dot{\varphi}+F_2\ddot{\varphi}+F_3\varphi^{(3)}\right)ds.
\eea
The corresponding equilibrium shape equation obtained by
$F_0-\frac{d F_1}{ds}+\frac{d^2F_2}{ds^2}-\frac{d^3F_3}{ds^3}=0$ is
\be
A(K-C_0)\dot{\tau}-(B-2A\tau)\dot{K}=0.
\ee
Eq. (32) can be changed into the following state:
\be
(K-C_0)^2(2\tau-Q)=C,
\ee
where $Q=B/A$, and $C$ is a constant. When $C_0=T_0=\lambda=0$,
similar equations are attained and an elliptic function solution is
found in Ref. [6]. In particular, if $\textbf{R}(s)$ is a planar
line ($\tau=B=T_0=0$), only Eq. (30) is valid and is reduced to an
elliptic different equation
\be
2A\ddot{K}+A K^3-(AC_0^2+2\lambda)K=0.
\ee
A similar equation was discussed in vesicle research.$^{27}$

Here, two special cases should be taken into account. First, there
is a ring solution to Eq. (34):
\be
K=1/r=\sqrt{C_0^2+2\lambda/A},
\ee
where $r$ is the the ring's radius. If there is stable planar ring
DNAs, we need
\be
 C_0^2+2\lambda/A>0.
\ee
Second, it might have isolated inflexion points in the two shape
equations at $s=s_0$, if
\be
K(s_0)=C_0.
\ee
When $C_0=0$, this problem is discussed in Ref. [6]. At the
inflexion points, DNA's shape may has uncertain behavior. Let's see
an example. Choosing $C=0$ in (33) and supposing $R(s)$ is a
solution of Eqs. (30) and (33), then, one can find that if there is
a point $s=s_0$ in $R(s)$ satisfies $K(s_0)=C_0$, Eq. (33) will lose
the constraint of $\tau$ and there is only Eq. (30) which needs to
be satisfied. In this case, Eq. (30) is easy to be satisfied,
because it has two variables: $K$ and $\tau$. Therefore, we can
expect that DNA's shape changes variably at its inflexion points.

\section*{3. General Shape Equations}

 It is known that the curvature $K$ and
torsion $\tau$ are two characteristic parameters which can determine
a line's shape. Recently, new energy terms are introduced$^{20-24}$
in DNA's total energy. Those new energy terms will influence DNA's
shape through changing its local curvature and tension. Moreover,
DNA's total energy can be written as the sum of the elastic energy
of its two molecular lines: $F_i=\oint [A K^2(s_i)+B \tau^2(s_i)]
ds$ ($i=1,2$) (see Ref. [4]). Where $s_i$ is the arclength of one of
DNA's molecular lines. Clearly, we can choose one of DNA's molecular
lines to determine its shape and attain its shape equations by
letting $\delta^{(1)} F_i=0$.

Here, we suppose that those new energy terms can be expressed by
curvature and tension, and we choose the total energy of closed DNA

\be
F=\oint f(K,\tau)ds,
\ee
where $f(K,\tau)$ is a function with two variables $K$ and $\tau$.
Similar method is adopted in vesicle research.$^{28,29}$ By the
former way we have discussed the first order variation of total
energy, that is
\be
\delta^{(1)} F=\oint [(f_{,K}\delta^{(1)} K+f_{,\tau}\delta^{(1)}
\tau)\sqrt{g}+f\delta^{(1)}(\sqrt{g})]dx,
\ee
where $f_{,K}=\frac{\partial f}{\partial K}$. Submitting (13), (21)
and (25) into (39), we get
\bea
\nonumber    F_0 &=& -K\psi,\\\nonumber
             F_1 &=& \left(K^2-\tau^2\right)\psi-\dot{\tau}\varphi-2\tau\dot{\varphi}+\ddot{\psi},\\\nonumber
             F_2 &=& K(2\tau\psi+\dot{\varphi})
                     +K^{-2}\dot{K}\left(\tau^2\varphi-\dot{\tau}\psi-2\tau\dot{\psi}-\ddot{\varphi}\right)\\ \nonumber
                 & & +K^{-1}\left[3\dot{\tau}\dot{\psi}-\tau^2\dot{\varphi}+\ddot{\tau}\psi+\varphi^{(3)}+
                     2\tau\left(\ddot{\psi}-\dot{\tau}\varphi\right)\right], \\
  \delta^{(1)} F &=& \oint \left(F_0 f+F_1f_{,K}+F_2f_{,\tau}\right)ds.
\eea
Perturbation only on main normal direction ($\varphi=0$) induces
\bea
\nonumber
               F_0 &=& \left(K^2-\tau^2\right)f_{,K}+K^{-2}\left(2K^3\tau-\dot{K}\dot{\tau}+K\ddot{\tau}\right)f_{,\tau}-K f,\\ \nonumber
               F_1 &=& K^{-2}(3K\dot{\tau}-2\dot{K}\tau)f_{,\tau},\\\nonumber
               F_2 &=& f_{,K}+2K^{-1}\tau f_{,\tau},\\
  \delta^{(1)} F   &=& \oint \left(F_0 \psi+F_1\dot{\psi}+F_2\ddot{\psi}\right)ds.
\eea
The corresponding equilibrium shape equation is
\bea
\nonumber
  \frac{d^2}{ds^2}(f_{,K}+2K^{-1}\tau f_{,\tau})-\frac{d}{ds}\left[K^{-2}(3K\dot{\tau}-2\dot{K}\tau)f_{,\tau}\right] & & \\
              +\left(K^2-\tau^2\right)f_{,K}+K^{-2}\left(2K^3\tau-\dot{K}\dot{\tau}+K\ddot{\tau}\right)f_{,\tau}-K f &=& 0.
\eea
If perturbation is on vice-normal direction ($\psi=0$), (40) is
simplified as
\bea
\nonumber
             F_0 &=& K^{-2}\tau(\dot{K}\tau-2K\dot{\tau})f_{,\tau}-\dot{\tau}f_{,K},\\\nonumber
             F_1 &=& K^{-1}\left(K^2-\tau^2\right)f_{,\tau}-2\tau f_{,K},\\\nonumber
             F_2 &=& -K^{-2}\dot{K}f_{,\tau},\\\nonumber
             F_3 &=& K^{-1}f_{,\tau},  \\
  \delta^{(1)} F &=& \oint \left(F_0 \varphi+F_1\dot{\varphi}+F_2\ddot{\varphi}+F_3\varphi^{(3)}\right)ds.
\eea
The corresponding equilibrium shape equation is
\bea
\nonumber
 \frac{d^3}{ds^3}\left(K^{-1}f_{,\tau}\right)+\frac{d^2}{ds^2}\left(K^{-2}\dot{K}f_{,\tau}\right)
                    +\frac{d}{ds}\left[K^{-1}\left(K^2-\tau^2\right)f_{,\tau}-2\tau f_{,K}\right] & & \\
                                  -K^{-2}\tau(\dot{K}\tau-2K\dot{\tau})f_{,\tau}+\dot{\tau}f_{,K} &=& 0.
\eea
 Eqs. (42) and (44) are the general equilibrium shape
equations for closed DNA. Comparing Eqs.(41) and (43) with Eqs.(6)
and (7) in Ref. [30] respectively, we find that Eq. (42) is equal to
Eq. (6) in Ref. [30] ($\alpha(s)=\alpha_s(s)=0$). But Eq. (43) is
not equal to Eq. (7) in Ref. [30]. Acutely, the mistakes in Ref.
[31] induce the incorrect result in Refs. [31] and [30] (see
appendix B). Moreover, similar correct equations are also shown in
Ref. [32] (Eqs.(16) and (17)), in Ref. [33] (Eqs.(1) and (2)) and in
Ref. [34] (Eqs.(77) and (78)).

\section*{4. Second Variation in planar case}

We know that stability condition needs the second variation
$\delta^{(2)}F>0$, so we want to get $\delta^{(2)}F$ by the former
way. For simplify, we only consider the planar case with the energy
$F=\oint (K^2+\lambda)ds$. Consider
$K^2=\ddot{\textbf{R}}\cdot\ddot{\textbf{R}}$, there is
\be
\delta{K^2}=2\ddot{\textbf{R}}\cdot\delta\ddot{\textbf{R}}+\delta\ddot{\textbf{R}}\cdot\delta\ddot{\textbf{R}}.
\ee
Submitting (19) into the above equation, we get
\bea
\nonumber
 \delta{K^2}&=& 2\ddot{\textbf{R}}\cdot\delta\ddot{\textbf{R}}+\delta\ddot{\textbf{R}}\cdot\delta\ddot{\textbf{R}}\\\nonumber
            &=& 2K(K^2\psi-\tau^2\psi-\dot{\tau}\varphi-2\tau\dot{\varphi}+\ddot{\psi})+3K^4\psi^2+\tau^4(\varphi^2+\psi^2)\\\nonumber
            & & +(\varphi^2+\psi^2)\dot{\tau}^2+4\tau^3(\psi\dot{\varphi}-\dot{\psi}\varphi)+2\dot{K}K\psi(\dot{\psi}-\tau\varphi)\\\nonumber
            & & +2\dot{\tau}(\psi\ddot{\varphi}-\ddot{\psi}\varphi)+\ddot{\psi}^2+\ddot{\varphi}^2-K^2\Big[\tau^2(\varphi^2+8\psi^2)+6\dot{\tau}\psi\varphi\\\nonumber
            & & +16\tau\psi\dot{\varphi}+2\dot{\varphi}^2-\dot{\psi}^2-6\ddot{\psi}\psi\Big]+2\tau^2(2\dot{\psi}^2+2\dot{\varphi}^2-\ddot{\psi}\psi-\ddot{\varphi}\varphi)\\
            & & +4\tau\Big[\dot{\tau}(\dot{\psi}\psi+\dot{\varphi}\varphi)+\dot{\psi}\ddot{\varphi}-\ddot{\psi}\dot{\varphi}\Big]+\emph{O(3)}
\eea
 Allying (13), we get the second variation of DNA's energy (note
$\tau=\varphi=0$)
\bea
\nonumber
\delta^{(2)}F &=& \oint\left[(K^2+\lambda)\delta^{(2)}\sqrt{g}+\sqrt{g}\delta^{(2)}K^2+\delta^{(1)}\sqrt{g}\delta^{(1)}K^2\right]dx\\
              &=& \oint\left[\frac{1}{2}(3K^2+\lambda)\dot{\psi}^2+4K^2\psi\ddot{\psi}+K^4\psi^2+2K\dot{K}\psi\dot{\psi}+\ddot{\psi}^2\right]ds.
\eea
The above equation can be used to analysis the stability of DNA in
planar case. For instance, it's easy to find that straight line
$K=0$ is a stable solution of Eq. (34) if $\lambda\geq0$.

\section*{5. Discussion}

In Sec. \textbf{2}, we give the DNA's shape equations which contain
two special constants: $C_0$ and $T_0$. But what are their physical
meaning? Maybe we can take them as the influence of the
environmental factors or other complications. For example, compared
with vesicle research, those two constants can be attributed to the
donation of electric potential on DNA.

In (47) we only give $\delta^{(2)}F$ in planar case, if we want to
get $\delta^{(2)}F$ in a more general case, $\delta^{(2)}K$ and
$\delta^{(2)}\tau$ should be obtained formerly. For instance, if we
consider
\bea
\delta X^n=nX^{n-1}\delta X+\frac{n(n-1)}{2!}(\delta X)^2+\cdots,
\eea
where $X=X(s)$ is a scalar function of $s$, and $n$ is a constant,
we can attain all $\delta K^n$ terms by using Eq. (46). Moreover,
using (16), all $\delta K^{(m)}$ ($K^{(m)}=\frac{d^m
K}{ds^m},~m=1,2\cdots$) terms can be obtained. Finally, because
attaining $\delta \tau^{(2)}$ needs a tedious calculation, we will
show it and discuss $\delta^{(2)}F$ in another paper.

\section*{Acknowledgements}
We would like to thank Jing Wang and Qing-hua Xu for their useful
discussions and suggestions.

\appendix{}
\section{}
  In this appendix, we show some useful identical equations which can
be attained easily.
\bea
 & & \dot{\textbf{n}}\cdot\dot{\textbf{R}}=-K,\\
 & & \dot{\textbf{n}}\cdot\ddot{\textbf{R}}=0,\\
 & & \dot{\textbf{n}}\cdot\textbf{R}^{(3)}=K^3+\tau^2K,\\
 & & \dot{\textbf{n}}\cdot\textbf{R}^{(4)}=3\dot{K}K^2+\tau(2\dot{K}\tau+K\dot{\tau}),\\
 & & \dot{\textbf{n}}\cdot \boldsymbol{\beta}=\tau,\\
 & & \dot{\textbf{n}}\cdot\dot{ \boldsymbol{\beta}}=0,\\
 & & \dot{\textbf{n}}\cdot\ddot{ \boldsymbol{\beta}}=-\tau^3-K^2\tau,\\
 & & \dot{\textbf{n}}\cdot \boldsymbol{\beta}^{(3)}=-K\dot{K}\tau-2K^2\dot{\tau}-3\tau^2\dot{\tau},\\
 & & \dot{\textbf{n}}\cdot\dot{\textbf{n}}=K^2+\tau^2,\\
 & & \dot{\textbf{n}}\cdot\ddot{\textbf{n}}=K\dot{K}+\tau\dot{\tau},\\
 & & \dot{\textbf{n}}\cdot\textbf{n}^{(3)}=K\ddot{K}+\tau\ddot{\tau}-(K^2+\tau^2)^2,\\
 & & \ddot{\textbf{n}}\cdot\ddot{\textbf{n}}=\dot{K}^2+\dot{\tau}^2+(K^2+\tau^2)^2.
\eea

\section{}
In this appendix, we point out a mistake in Ref. [31]. Comparing
this letter with Ref. [31], we find that they are similar work. Here
we use $\psi$ and $\varphi$ to determine the perturbations on the
main normal direction $\textbf{n}$ and the the vice-normal direction
 $\boldsymbol{\beta}$, respectively. (Actually, there should be
perturbation on the tangent direction $\boldsymbol{\alpha}$, but we
find that it will not induces any new shape equation, so we ignore
it deliberately.) In Ref. [31], there are three corresponding
perturbation functions: $\varepsilon_1$ on the tangent direction
$\boldsymbol{\alpha}$, $\varepsilon_2$ on the main normal direction
$\textbf{n}$ and $\varepsilon_3$ on the vice-normal direction
 $\boldsymbol{\beta}$. Clearly, there are the relationships
\be
\psi=\varepsilon_2,~\varphi=\varepsilon_3.
\ee
Comparing Eq. (25) with Eq. (2.27) in Ref. [31], we find they are
not equal. There are two mistakes in Eq. (2.27) in Ref. [31] that
the first term $K^2\varepsilon_3$ in the first bracket should be
changed as $\tau^2\varepsilon_3$, and the $-\varepsilon_3^{(3)}$
term should be $\varepsilon_3^{(3)}$. The correct $\delta^{(1)}\tau$
in Eq. (2.27) in Ref. [31] should be
\bea
\nonumber
\delta^{(1)}\tau &=& \frac{1}{K^2}\bigg\{\dot{K}(\tau^2\varepsilon_3-\dot{\tau}\varepsilon_2-2\tau\dot{\varepsilon}_2-\ddot{\varepsilon}_3)\\
                 & & +K\left[2K^2\tau\varepsilon_2+\dot{\tau}(-2\tau\varepsilon_3+3\dot{\varepsilon}_2)
                        +(K^2-\tau^2)\dot{\varepsilon}_3+\ddot{\tau}\varepsilon_2+2\tau\ddot{\varepsilon}_2+\varepsilon_3^{(3)}\right]\bigg\}.
\eea
One can find that the above equation is equal to Eq. (25) (note
$\varepsilon_1=0$). It is this mistake that induces the incorrect
shape equations: Eq. (2.32) in Ref. [31] and Eq. (7) in Ref. [30].

\end{document}